%% file: paper.tex
\documentclass[conference]{IEEEtran}
\IEEEoverridecommandlockouts
\pdfoutput=1	
\usepackage{booktabs} 
\usepackage{cite}

\usepackage{graphicx}
\usepackage[utf8]{inputenc}
\usepackage{algorithm}
\usepackage[noend]{algpseudocode}
\usepackage{enumitem}
\graphicspath{ {diagrams/CA-CQR2/} {ca-cqr2plots/} }

\usepackage{xcolor}    
\definecolor{darkGreen}{RGB}{0,153,0}

\usepackage{makecell}
\usepackage{indentfirst}		

\usepackage{epsfig}  
\usepackage{subfigure}  
\usepackage[cmex10]{amsmath}  
\usepackage{amssymb}  
\usepackage{url}  
\usepackage{lscape}  
\usepackage[justification=raggedright]{caption}	

\usepackage[english]{babel}
\usepackage{verbatim}             
\usepackage[margin=0.75in]{geometry} 
\usepackage{ulem}
\usepackage {multirow}
\usepackage{capt-of}		  

\usepackage{amsfonts}

\newcommand{\algsmall}{}
\newcommand{\mathsmall}{}
\newcommand{\CholInv}{\textbf{CholInv}}
\newcommand{\CFRDDD}{\textbf{CFR3D}}
\newcommand{\fleft}[1]{\big#1}
\newcommand{\fright}[1]{\big#1}
\newcommand{\sfrac}[2]{{#1}/{#2}}
\newcommand{\ttransp}[2]{T_{\text{Transp}}^{\alpha-\beta}\big(#1,#2\big)}
\newcommand{\tbcast}[2]{T_{\text{Bcast}}^{\alpha-\beta}\big(#1,#2\big)}
\newcommand{\tred}[2]{T_{\text{Reduce}}^{\alpha-\beta}\big(#1,#2\big)}
\newcommand{\tallred}[2]{T_{\text{Allreduce}}^{\alpha-\beta}\big(#1,#2\big)}
\newcommand{\tallgat}[2]{T_{\text{Allgather}}^{\alpha-\beta}\big(#1,#2\big)}
\newcommand{\tmm}[3]{T_{\text{MM}}^{\alpha-\beta}\big(#1,#2,#3\big)}
\newcommand{\taxpy}[2]{T_{\text{axpy}}^{\alpha-\beta}\big(#1,#2\big)}
\newcommand{\tsyrk}[2]{T_{\text{syrk}}^{\alpha-\beta}\big(#1,#2\big)}
\newcommand{\tcholinv}[1]{T_{\text{CholInv}}^{\alpha-\beta}\big(#1\big)}
\newcommand{\tchol}[1]{T_{\text{Chol}}^{\alpha-\beta}\big(#1\big)}
\newcommand{\tcfrddd}[2]{T_{\text{CFR3D}}^{\alpha-\beta}\big(#1,#2\big)}
\newcommand{\tmmddd}[4]{T_{\text{MM3D}}^{\alpha-\beta}\big(#1,#2,#3,#4\big)}
\newcommand{\tdcqr}[3]{T_{\text{1D-CQR}}^{\alpha-\beta}\big(#1,#2,#3\big)}

\newcommand{\tcacqr}[4]{T_{\text{CA-CQR}}^{\alpha-\beta}\big(#1,#2,#3,#4\big)}
\newcommand{\px}{\mathbf{x}}
\newcommand{\py}{\mathbf{y}}
\newcommand{\pz}{\mathbf{z}}
\newcommand{\Id}[1]{I}
\newcommand{\Zr}[1]{0}
\newcommand{\loc}[1]{\Pi\langle{#1}\rangle}

\def \ALL {{\textbf{:}}}

\begin{document}

\title{Communication-avoiding CholeskyQR2 for rectangular matrices
}

\author{\IEEEauthorblockN{Edward Hutter}
\IEEEauthorblockA{\textit{Department of Computer Science} \\
\textit{University of Illinois at Urbana-Champaign}\\
hutter2@illinois.edu}
\and
\IEEEauthorblockN{Edgar Solomonik}
\IEEEauthorblockA{\textit{Department of Computer Science} \\
\textit{University of Illinois at Urbana-Champaign}\\
solomon2@illinois.edu}
}

\maketitle

\begin{abstract}
\input{abs}
\end{abstract}

\input{body}

\bibliographystyle{IEEEtran}
\bibliography{paper}

\end{document}

%% file: abs.tex
\par{}
Scalable QR factorization algorithms for solving least squares and eigenvalue problems are critical given the increasing parallelism within modern machines. We introduce a more general parallelization of the CholeskyQR2 algorithm and show its effectiveness for a wide range of matrix sizes. Our algorithm executes over a 3D processor grid, the dimensions of which can be tuned to trade-off costs in synchronization, interprocessor communication, computational work, and memory footprint. We implement this algorithm, yielding a code that can achieve a factor of $\Theta(P^{1/6})$ less interprocessor communication on $P$ processors than any previous parallel QR implementation. Our performance study on Intel Knights-Landing and Cray XE supercomputers demonstrates the effectiveness of this CholeskyQR2 parallelization on a large number of nodes. Specifically, relative to ScaLAPACK's QR, on 1024 nodes of Stampede2, our CholeskyQR2 implementation is faster by 2.6x-3.3x in strong scaling tests and by 1.1x-1.9x in weak scaling tests.

%% file: body.tex
\section{Introduction}

\par{}
The reduced QR factorization $A=QR$, $Q\in\mathbb{R}^{m\times n}$ with orthonormal columns and upper-triangular $R\in\mathbb{R}^{n\times n}$, is a ubiquitous subproblem in numerical algorithms. It can be used to solve linear systems, least squares problems, as well as eigenvalue problems with dense and sparse matrices. In many application contexts, these problems involve very overdetermined systems of equations in a large number of variables, driving a need for scalable parallel QR factorization algorithms. We study communication-efficient algorithms for QR factorizations of rectangular matrices $m\geq n$.

\par{}
Our work builds on a recently developed algorithm, CholeskyQR2\cite{Fukaya:2014:CSC:2691142.2691147}, a Cholesky-based algorithm designed for tall and skinny matrices. The well-known CholeskyQR algorithm computes $\hat{R}\approx R$ as the upper triangular factor in the Cholesky factorization of $n\times n$ matrix $A^TA$, then solves $m$ triangular linear systems of equations, $\hat{Q} = A\hat{R}^{-1}$. The forward error in computing $\hat{Q}$ in CholeskyQR can be as high as $\Theta(\kappa(A)^2)$ where $\kappa(A)$ is the condition number of $A$. On the other hand, the triangular solves are backward stable, so $\hat{Q}\hat{R}=A+\delta A$ with $||\delta A||=O(\epsilon)$ where $\epsilon$ is the relative error in the machine floating point representation. Therefore, when under effect of round-off error, CholeskyQR computes an accurate factorization of $A$ into a product of a nearly orthogonal matrix $\hat{Q}$ and a triangular matrix $\hat{R}$.

\par{}
\begin{table}[h]
\renewcommand{\arraystretch}{1.5}
\begin{center}
\begin{tabular}{r|c|c|c}
{\bf algorithm} & {\bf latency} ($\alpha$) & {\bf bandwidth} ($\beta$) & {\bf flops} ($\gamma$) 
 \\
\hline
{\bf MM3D} & $\log P$ & $\frac{mn+nk+mk}{P^{2/3}}$ & $mnk/P$  
 \\
\hline
{\bf CFR3D} & $P^{2/3} \log P$ & $n^2/P^{2/3}$ & $n^3/P$  
 \\
\hline
{\bf 1D-CQR} & $\log P$ & $n^2$ & $mn^2/P+n^3$  
 \\
\hline
{\bf 3D-CQR} & $P^{2/3} \log P$ & $mn/P^{2/3}$ & $mn^2/P$  
 \\
\hline
{\bf CA-CQR} & $c^2 \log P$ & $\frac{mn}{dc} +\frac{n^2}{c^2}$ & $ \frac{mn^2}{P}+\frac{n^3}{c^3}$  
 \\
\hline
{\bf CA-CQR} & $(Pn/m)^{2/3}\log P$ & $(mn^2/P)^{2/3}$ & $ mn^2/P$ \\ 
\end{tabular}
\end{center}
\caption{Summary of asymptotic costs for each algorithm described in the paper for QR of $m\times n$ matrix on $P$ processors (for CA-CQR first using $c\times d\times c$ processor grid, then with best choice of $c,d$). CA-CQR2 achieves same asymptotic costs as CA-CQR.}
\label{tab:allcosts}
\end{table}


\par{}
CholeskyQR2 provides a correction to the orthogonal factor by running CholeskyQR once more on $\hat{Q}$ itself to obtain $Q {\delta R} = \hat{Q}$, where ${\delta R}$ is upper-triangular and near identity. The upper-triangular factor can then be updated as $R={\delta}R \hat{R}$. If $\hat{Q}$ was computed to within a few digits of accuracy, which is guaranteed if $\kappa(A)=O(\sqrt{1/\epsilon})$, it will be close to orthogonal and therefore well-conditioned. In this case, the CholeskyQR of $\hat{Q}$ will not lose much precision from round-off error and the QR factorization given by CholeskyQR2 will be as accurate as Householder QR \cite{yamamoto2015roundoff}.
Recently, an extension to CholeskyQR2 has been proposed that uses a third CholeksyQR execution to achieve unconditional stability (as good as Householder QR)~\cite{2018arXiv180911085F}.

\par{}
In previous work \cite{BDGJNS_IPDPS_2014}, a 1D parallelization of the algorithm was shown to provide minimal communication cost and synchronization (a logarithmic factor less than other communication-avoiding algorithms \cite{Demmel:2012:CPS:2340316.2340324}). In this work, we extend the algorithm to a 3D parallelization that requires minimal communication cost for matrices with arbitrary dimensions. 
We utilize efficient parallel algorithms for Cholesky factorization and matrix multiplication. Communication-efficient of matrix multiplication and Gaussian elimination algorithms are well-studied and have demonstrated performance improvement in implementation. We present a simple adaptation of the algorithm therein \cite{Tiskin2002} to Cholesky factorization. Tradeoffs between synchronization and bandwidth are investigated using a tunable parameter, $\phi$, representing the depth of recursion. 
We focus on minimizing bandwidth cost given unbounded memory. We provide a summary of the cost analysis of the algorithm and its main components in Table~\ref{tab:allcosts}.

\begin{figure*}[t]
\centering
\subfigure[Strong scaling for matrices with dimensions given in legend.]{
\includegraphics[width=0.48\textwidth]{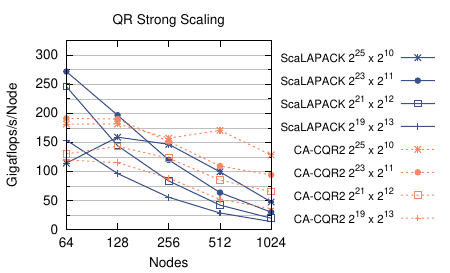}
\label{fig:BestSSvariants}
}
\subfigure[Weak scaling for $m\times n$ matrices so $mn^2$ scales linearly with node count.]{
\includegraphics[width=0.48\textwidth]{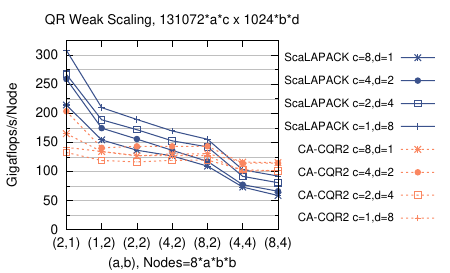}
\label{fig:BestWSvariants}
}
\caption{Performance on Stampede2 with 64 MPI processes/node of CholeskyQR2 and ScaLAPACK.}
\end{figure*}

\par{}A key advantage of CholeskyQR2 is its practicality. It requires only matrix multiplications and Cholesky factorizations. 
Of the QR algorithms that achieve asymptotically minimal communication (match 3D matrix multiplication in bandwidth cost)~\cite{tiskin2007communication, Solomonik:2017:CPA:3087556.3087561,ES_dissertation_2014,BDGJK18}, none have been implemented in practice. 
Our implementation of the 3D Cholesky-QR2 algorithm  (CA-CQR2) shows that the improvement in communication cost allows improvements over the performance of a state-of-the-art library for parallel QR (ScaLAPACK's PGEQRF) on large node counts. Our performance results are summarized in Figure~\ref{fig:BestSSvariants} for strong scaling and Figure~\ref{fig:BestWSvariants} for weak scaling, which show performance for the median execution time of 5 iterations for the best performing choice of processor grid at each node count.


\par{}
3D QR factorization algorithms face challenges in their complexity, high constants, and incompatibility with BLAS-like routines. Previously studied parallel CholeskyQR2 (CQR2) algorithms do not scale for matrices of an arbitrary size. The merit of our communication-avoiding CholeskyQR2 (CA-CQR2) algorithm lies in the combination of its simplicity and asymptotic efficiency. It relies upon a tunable processor grid to ensure optimally minimal communication for matrices of any size while being easy to implement. Our contributions are a detailed specification, cost analysis, implementation, and performance evaluation of this algorithm.

\section{Foundation and Previous Work}
\par{}
We study the scalability of parallel QR factorization algorithms via interprocessor communication cost analysis. In this section, we introduce results necessary for our work, and provide brief summaries of related work. We first consider collective communication and the parallel algorithms we use for matrix multiplication and Cholesky factorization. 

\subsection{Preliminaries}
In presenting these algorithms and analyzing their costs, we use the $\alpha$-$\beta$-$\gamma$ model as defined below.
{\mathsmall
\begin{equation*}
\begin{aligned}
\alpha &\rightarrow \text{cost of sending or receiving a single message}\\
\beta &\rightarrow \text{cost of moving a single word of data among processors}\\
\gamma &\rightarrow \text{cost of computing a single floating point operation}\\
\end{aligned}
\end{equation*}
}%
Our analysis assumes $\alpha \gg \beta \gg \gamma$, which is reflective of current parallel computer architectures.
\par{}
We define a few sequential routines and give their asymptotic costs~\cite{lawson1979basic}.
{\mathsmall
\begin{alignat*}{3}
C \gets aX+Y \  : \  & \taxpy{m}{n} && =\  &&2mn \cdot \gamma \\
C \gets \textbf{MM}(A,B)=AB \  : \  & \tmm{m}{n}{k} &&=\  &&2mnk \cdot \gamma \\
C \gets \textbf{Syrk}(A)=A^TA \  : \  & \tsyrk{m}{n} &&=\  &&mn^{2} \cdot \gamma \\
L \gets \textbf{Chol}(A)=LL^T \  : \  & \tchol{n} &&=\  &&(2n^{3}/3) \cdot \gamma 
\end{alignat*}
}%

\subsection{Processor Grids and Collective Communication}
\par{}
Collective communication serves as an efficient way to move data among processors over some subset of a processor grid. We define a 3D processor grid $\Pi$ containing $P$ processors. $\Pi[\px,\py,\pz]$ uniquely identifies every processor in the grid, where each of the corresponding dimensions are of size $P^{\frac{1}{3}}$ and $\px,\py,\pz \in [0,P^{\frac{1}{3}}{-}1]$. $\Pi$ can be split into 2D slices such as $\Pi[\ALL,\ALL,\pz]$, row communicators such as $\Pi[\ALL,\py,\pz]$, column communicators such as $\Pi[\px,\ALL,\pz]$, and depth communicators such as $\Pi[\px,\py,\ALL]$.

\par{}
Allgather, Allreduce, and Broadcast are collectives used heavily in the multiplication and factorization algorithms explored below. As these are well known~\cite{geijn,rec_double_and_half,bruck_algo,SAAD1989115,traff2008optimal}, we give only the function signature and a brief description of each. We assume butterfly network collectives for cost analysis~\cite{rec_double_and_half}, which are optimal or close to optimal in the $\alpha$-$\beta$-$\gamma$ model.
\begin{itemize}
\item \textbf{Transpose}$\left(A, \Pi[\py,\px,\pz]\right)$ : all processors $\Pi[\px,\py,\pz]$ swap local array $A$ with processor $\Pi[\py,\px,\pz]$ via point-to-point communication.
\item \textbf{Bcast}$\left(A, B, r, \Pi[\ALL,\py,\pz]\right)$ : root processor $\Pi[r,\py,\pz]$ distributes local array $A$ to every processor in $\Pi[\ALL,\py,\pz]$ as local array $B$.
\item \textbf{Reduce}$\left(A, B, r, \Pi[\px,\ALL,\pz]\right)$ : all processors in $\Pi[\px,\ALL,\pz]$ contribute local arrays $A$ to an element-wise reduction onto root processor $\Pi[\px,r,\pz]$ as local array $B$.
\item \textbf{Allreduce}$\left(A, B, \Pi[\px,\ALL,\pz]\right)$ : all processors in $\Pi[\px,\ALL,\pz]$ contribute local arrays $A$ to an element-wise reduction, the result of which is broadcasted into local array $B$.
\item \textbf{Allgather}$\left(A, B, \Pi[\px,\py,\ALL]\right)$ : all processors in $\Pi[\px,\py,\ALL]$ contribute local arrays $A$ to a concatenation, the result of which is broadcasted into local array $B$.
\end{itemize}

\par{}
The costs of these collective routines can be obtained by a butterfly schedule, where $n$ words of data are being communicated across $P$ processors.
{\mathsmall
\begin{equation*}
\begin{aligned}
\ttransp{n}{P} &= \delta(P)\left(\alpha + n\cdot\beta\right) \\
\tbcast{n}{P}&= 2\log_{2} P\cdot\alpha + 2n\delta(P)\cdot\beta \\
\tred{n}{P} &= 2\log_{2} P\cdot\alpha + 2n\delta(P)\cdot\beta \\
\tallred{n}{P} &= 2\log_{2} P\cdot\alpha + 2n\delta(P)\cdot\beta \\
\tallgat{n}{P} &= \log_{2} P\cdot\alpha + n\delta(P)\cdot\beta \\
\end{aligned}
\end{equation*}
}%
\par{}
where
{\mathsmall
\begin{equation*}
  \delta(x) = \left\{
  \begin{array}{ll}
  0 & \quad x \leq 1 \\
  1 & \quad x > 1
  \end{array}
  \right\}.
\end{equation*}
}%
We disregard the computational cost in reductions by the assumption $\beta \gg \gamma$.

\subsection{Matrix Multiplication}
\par{}
Matrix multiplication $C=AB$ over $\Pi$ and other cubic partitions of $\Pi$ is an important building block for the 3D-CQR2 and CA-CQR2 algorithms presented below. We use a variant of 3D matrix multiplication (which we refer to as 3D SUMMA) that achieves asymptotically optimal communication cost over a 3D processor grid \cite{mccol_tiskin_99,dekel:657,matmul3d,snirmatmul,berntsen1989communication,Johnsson:1993:MCT:176639.176642}. Our algorithm MM3D is a customization with respect to known algorithms. First, $B$ is not distributed across $\Pi[P^{\frac{1}{3}}{-}1,\ALL,\ALL]$ and is instead distributed across $\Pi[\ALL,\ALL,0]$ with $A$. Second, instead of distributing matrix $C$ across $\Pi[\ALL,0,\ALL]$, we Allreduce $C$ onto $\Pi[\ALL,\ALL,\pz], \forall \pz \in [0,P^{\frac{1}{3}}{-}1]$ so that each 2D slice holds a distributed copy. These differences are motivated by the need for $C$ to be replicated over each 2D slice of $\Pi$ in our new algorithms.
A cyclic distribution is used as it is required by our 3D Cholesky factorization algorithm detailed below. See Algorithm~\ref{alg:mm3d} for specific details.

\par{}
\begin{algorithm}[t]
{\algsmall
\caption {$[\loc{C}]\gets \textbf{MM3D}\fleft(\loc{A}, \loc{B}, \Pi\fright)$}
\label{alg:mm3d}
\begin{algorithmic}[1]
\Require{$\Pi$ has $P$ processors arranged in a $P^{1/3}\times P^{1/3}\times P^{1/3}$ grid. Matrices $A$ and $B$ are replicated on $\Pi[\ALL,\ALL,\pz], \forall \pz \in [0,P^{\frac{1}{3}}{-}1]$. Each processor $\Pi[\px,\py,\pz]$ owns a cyclic partition of $m\times n$ matrix $A$ and $n\times k$ matrix $B$, referred to as local matrices $\loc{A}$ and $\loc{B}$, respectively.
Let $X$, $Y$, and $Z$ be temporary arrays with the same distribution as $A$ and $B$.}
  \State \textbf{Bcast}$\fleft(\loc{A}, \loc{X}, \pz, \Pi[\ALL,\py,\pz]\fright)$ \label{li:MM3D:bcast1}
  \State \textbf{Bcast}$\fleft(\loc{B}, \loc{Y}, \pz, \Pi[\px,\ALL,\pz]\fright)$ \label{li:MM3D:bcast2}
  \State $\loc{Z}\gets $\textbf{MM}$\fleft(\loc{X}, \loc{Y}\fright)$ \label{li:MM3D:loc}
  \State \textbf{Allreduce}$\fleft(\loc{Z}, \loc{C}, \Pi[\px,\py,\ALL]\fright)$ \label{li:MM3D:allred}
\Ensure{$C=AB$, where $C$ is $m\times k$ and distributed the same way as $A$ and $B$.}
\end{algorithmic}
}
\end{algorithm}

\par{}
The cost of MM3D for multiplying an $m\times n$ matrix by a $n\times k$ matrix is given in Table \ref{tab:allcosts}.

\subsection{Cholesky Factorization}
\par{}
Assuming $A$ is a dense, symmetric positive definite matrix of dimension $n$, the factorization $A = LL^{T}$ can be expanded into matrix multiplication of submatrices of dimension $\sfrac{n}{2}$ \cite{gustavson1997recursion,Tiskin2002},
{\mathsmall
\begin{equation*}
\begin{aligned}
&\begin{bmatrix}
A_{11} \\
A_{21} & A_{22}\end{bmatrix}
 = \begin{bmatrix}
	L_{11} \\
	L_{21} & L_{22} \end{bmatrix}
\begin{bmatrix}
	L^{T}_{11} & L^{T}_{21} \\
	& L^{T}_{22} \end{bmatrix},
\end{aligned}
\end{equation*}
}%
{\mathsmall
\begin{equation*}
\begin{aligned}
A_{11}&=L_{11}L^{T}_{11}, \quad 
A_{21}=L_{21}L^{T}_{11}, \\
A_{22}&=L_{21}L^{T}_{21}+L_{22}L^{T}_{22}.
\end{aligned}
\end{equation*}
}%
\par{}
This recursive algorithm yields a family of parallel algorithm variants~\cite{george1986parallel}.
Rewriting these equations gives a recursive definition for $L \gets \textbf{Chol}\fleft(A\fright)$,
{\mathsmall
\begin{equation*}
\begin{aligned}
L_{11} &\gets \textbf{Chol}\fleft(A_{11}\fright),
\quad \quad L_{21} \gets A_{21}L_{11}^{-T}, \\
L_{22} &\gets \textbf{Chol}\fleft(A_{22}{-}L_{21}L^{T}_{21}\fright).
\end{aligned}
\end{equation*}
}%
\par{}
We augment the recursive definition of the Cholesky factorization, using the recursive definition for $Y=L^{-1}$, as motivated by other parallel algorithms leveraging the triangular inverse~\cite{7967158,Tiskin2002}. Like before, the factorization $\Id{n} = LY$ gets expanded into matrix multiplication of submatrices of dimension $\sfrac{n}{2}$,
\par{}
{\mathsmall
\begin{equation*}
\begin{aligned}
\begin{bmatrix}
\Id{\frac{n}{2}} & 0 \\
0 & \Id{\frac{n}{2}}
\end{bmatrix}
&=
\begin{bmatrix}
L_{11} & \\
L_{21} & L_{22}
\end{bmatrix}
\begin{bmatrix}
Y_{11} \\
Y_{21} & Y_{22}
\end{bmatrix},
\end{aligned}
\end{equation*}
\begin{equation*}
\begin{aligned}
\Id{\frac{n}{2}} = L_{11}Y_{11}, \quad
\Id{\frac{n}{2}} = L_{22}Y_{22}, \quad
\Zr{\frac{n}{2}} = L_{21}Y_{11} + L_{22}Y_{21}.
\end{aligned}
\end{equation*}
}%
\par{}
Rewriting these equations gives a recursive definition for $Y \gets \textbf{Inv}\left(L\right)$,
{\mathsmall
\begin{equation*}
\begin{aligned}
Y_{11} \gets \textbf{Inv}\left(L_{11}\right), \ 
Y_{22} \gets \textbf{Inv}\left(L_{22}\right), \ 
Y_{21} \gets -Y_{22}L_{21}Y_{11}.
\end{aligned}
\end{equation*}
}%

\par{}
We embed the two recursive definitions and arrive at algorithm \ref{alg:cfrseq}, which solves both the Cholesky factorization of $A$ and the triangular inverse of $L$. Note that the addition of solving for $L^{-1}$ adds only two extra matrix multiplications at each recursive level to the recursive definition for $A=LL^{T}$, thus achieving the same asymptotic cost. If $L^{-1}$ were to be solved recursively at each level, the synchronization cost would incur an extra logarithmic factor. We address the missing base case in the cost analysis derivation below.

\par{}
\begin{algorithm}[h]
{\algsmall
\caption{$[L, Y]\gets$\CholInv$\left(A\right)$}
\label{alg:cfrseq}
\begin{algorithmic}[1]
\Require{$A$ is a symmetric positive definite matrix of dimension $n$.}
  \State if $n=1$ return $L=\text{Chol}(A),Y=L^{-1}$
  \State 
\(
L_{11} , Y_{11}
\gets \CholInv\left(A_{11}\right) 
\)
  \State \(
L_{21} \gets A_{21}Y_{11}^{T} 
\)
  \State
\(L_{22} , Y_{22}
\gets \CholInv\left(A_{22}-L_{21}L^{T}_{21}\right)
\)
  \State 
\(L \gets
\begin{bmatrix}
L_{11} & \\
L_{21} & L_{22}
\end{bmatrix}
\)
\State
\( 
Y \gets
\begin{bmatrix}
Y_{11} & \\
-Y_{22}L_{21}Y_{11} & Y_{22}
\end{bmatrix}
\)
\State return $L,Y$
\Require{$L$ is lower triangular, $A=LL^T$.}

\end{algorithmic}
}%
\end{algorithm}

\par{}
We incorporate two matrix transposes at each recursive level to take into account $L_{11}^{-T}$ and $L_{21}^{T}$ needed in the equations above. Processor $\Pi[\px,\py,\pz], \forall \pz \in [0,P^{\frac{1}{3}}{-}1]$ must send its local data to $\Pi[\py,\px,\pz]$ to transpose the matrix globally.

\par{}
A cyclic distribution of the matrices among processors $\Pi[\ALL,\ALL,\pz], \forall \pz \in [0,P^{\frac{1}{3}}{-}1]$ is chosen to utilize every processor in the recursive calls performed on submatrices. Upon reaching the base case where matrix dimension $n=n_{o}$, the submatrix is scattered over the $P^{\frac{2}{3}}$ processors in $\Pi[\ALL,\ALL,\pz], \forall \pz \in [0,P^{\frac{1}{3}}{-}1]$ and an Allgather is performed so that all processors obtain the submatrix. Then all $P$ processors perform a Cholesky factorization and triangular inverse redundantly. See Algorithm~\ref{alg:cfrddd} for full details.

\par{}
\begin{algorithm}[h]
{\algsmall
\caption {$[\loc L,\loc Y] \gets \CFRDDD\left({\loc A, n_{o}, \Pi}\right)$}
\label{alg:cfrddd}
\begin{algorithmic}[1]
\Require{$\Pi$ has $P$ processors arranged in a $P^{1/3}\times P^{1/3}\times P^{1/3}$ 3D grid. Symmetric positive definite matrix $A\in\mathbb{R}^{n\times n}$ is replicated on $\Pi[\ALL,\ALL,\pz], \forall \pz \in [0,P^{\frac{1}{3}}{-}1]$. Each processor $\Pi[\px,\py,\pz]$ owns a cyclic partition of $A$ given by $\loc{A}$.
Let $T$, $W$, $X$, $U$, and $Z$ be temporary arrays, distributed the same way as $A$.}
  \If {$n = n_{o}$}
    \State $\textbf{Allgather}\fleft(\loc{A}, T, \Pi[\ALL,\ALL,\pz]\fright)$ \label{li:cfr3d:ballg}
    \State $L,Y \gets \textbf{CholInv}\fleft(T, n\fright)$ \label{li:cfr3d:cholinv} \Comment{$\loc{L},\loc{Y} \text{ are saved}$}
  \Else
  \State $\loc{L_{11}}, \loc{Y_{11}} \gets \CFRDDD\fleft(\loc{A_{11}}, n_{o}, \Pi\fright)$ \label{li:cfr3d:rec1}
  \State $\loc{W} \gets \textbf{Transpose}\fleft(\loc{Y_{11}}, \Pi[\py,\px,\pz]\fright)$ \label{li:cfr3d:tr1}
  \State $\loc{L_{21}} \gets \textbf{MM3D}\fleft(\loc{A_{21}}, \loc{W}^{T}, \Pi\fright)$ \label{li:cfr3d:mm3}
  \State $\loc{X} \gets \textbf{Transpose}\fleft(\loc{L_{21}}, \Pi[\py,\px,\pz]\fright)$ \label{li:cfr3d:tr2}
  \State $\loc{U} \gets \textbf{MM3D}\fleft(\loc{L_{21}}, \loc{X}^{T}, \Pi\fright)$ \label{li:cfr3d:mm4}
  \State $\loc{Z} \gets \loc{A_{22}} - \loc{U}$ \label{li:cfr3d:sub}
  \State $\loc{L_{22}}, \loc{Y_{22}} \gets \CFRDDD\fleft(\loc{Z}, n_{o}, \Pi\fright)$ \label{li:cfr3d:rec2}
  \State $\loc{U} \gets \textbf{MM3D}\fleft(\loc{L_{21}}, \loc{Y_{11}}, \Pi\fright)$ \label{li:cfr3d:mm5}
  \State $\loc{W} \gets -\loc{Y_{22}}$ \label{li:cfr3d:neg}
  \State $\loc{Y_{21}} \gets \textbf{MM3D}\fleft(\loc{W}, \loc{U}, \Pi\fright)$ \label{li:cfr3d:mm6}
  \EndIf
\Ensure{$A=LL^{T}$, $Y=L^{-1}$, where matrices $L$ and $Y$ are distributed the same way as $A$.}
\end{algorithmic}
}
\end{algorithm}

\par{}
The combined communication cost of the base case does not dominate the cost of the MM3D. We analyze the cost of CFR3D in Table \ref{table:CFcosts}.

\par{}
{\mathsmall
\begin{table}[h]
\begin{center}
\renewcommand{\arraystretch}{1.5}
\begin{tabular}{ l | l }
\# & Cost \\ \hline
\ref{li:cfr3d:ballg}& $\tallgat{n_o^2}{P^{2/3}}$ 
 \\ \hline 
\ref{li:cfr3d:cholinv} & $\tcholinv{n_o}$ 
 \\ \hline 
\ref{li:cfr3d:rec1} & $\tcfrddd{n/2}{P}$ 
 \\ \hline 
\ref{li:cfr3d:tr1} & $\ttransp{n^2/(8P^{2/3})}{P^{2/3}}$ 
 \\ \hline 
\ref{li:cfr3d:mm3} & $\tmmddd{n/2}{n/2}{n/2}{P}$ 
 \\ \hline 
\ref{li:cfr3d:tr2} & $\ttransp{n^2/(4P^{2/3})}{P^{2/3}}$ 
 \\ \hline 
\ref{li:cfr3d:mm4} &  $\tmmddd{n/2}{n/2}{n/2}{P}$ 
 \\ \hline 
\ref{li:cfr3d:sub} & $\taxpy{n/2}{n/2}$ 
 \\ \hline 
\ref{li:cfr3d:rec2} & $\tcfrddd{n/2}{P}$
 \\ \hline 
\ref{li:cfr3d:mm5} & $\tmmddd{n/2}{n/2}{n/2}{P}$ 
 \\ \hline 
\ref{li:cfr3d:mm6} & $\tmmddd{n/2}{n/2}{n/2}{P}$ 
 \\ 
\end{tabular}
\end{center}
\caption{Per-line costs of Algorithm \ref{alg:cfrddd}.}
\label{table:CFcosts}
\end{table}
}%

{\mathsmall
\begin{equation*}
\begin{aligned}
T_{\text{BaseCase}}^{\alpha-\beta}\left(n_{o},P\right) &= \frac 23\log_{2} P \cdot\alpha + n_{o}^{2}\delta(P) \cdot \beta + \frac 12 n_{o}^{3} \cdot \gamma \\
&= \mathcal{O}{\left(\log P \cdot \alpha + n_{o}^{2}\delta(P) \cdot \beta + n_{o}^{3} \cdot \gamma\right)} \\
\end{aligned}
\end{equation*}
}%
\par{}
Choice of $n_{o}$ depends on the non-recursive communication cost. Our algorithm computes $\frac{n}{n_{o}}$ Allgathers, one for each base case. The total cost is as follows:
{\mathsmall
\begin{equation*}
\begin{aligned}
&\tcfrddd{n}{P} = 2\tcfrddd{n/2}{P} + \left(8\log_{2} P+2\delta(P)\right) \cdot \alpha \\
& \hspace{2em} + \frac{\left(45n^{2}+14n\right)\delta(P)}{8P^{\sfrac{2}{3}}} \cdot \beta + \frac{n^{3}}{P} \cdot \gamma \\
&= \mathcal{O}\Big(\frac{n\log P}{n_{o}} \cdot \alpha + n\big(n_{o}{+}\frac{n}{P^{\sfrac{2}{3}}}\big) \cdot \beta + n\big(n_{o}^{2}{+}\frac{n^2}{P}\big) \cdot \gamma\Big).
\end{aligned}
\end{equation*}
}%

\par{}
Choice of $\frac{n}{n_{o}}$ creates a tradeoff between the synchronization cost and the communication cost. We minimize communication cost over synchronization by choosing $n_{o} = n/P^{\frac{2}{3}}$. The resulting cost of the 3D algorithm is listed in Table~\ref{tab:allcosts}.

\subsection{QR Factorization}
\par{}
QR factorization decomposes an $m\times n$ matrix $A$ into matrices $Q$ and $R$ such that $A=QR$. We focus on the case when $m\geq n$ and $Q$ and $R$ are the results of a reduced QR factorization. In this case, $Q$ is $m\times n$ with orthonormal columns and $R$ is $n\times n$ and upper-triangular. 
Parallel QR algorithms have received much study~\cite{o1990parallel,elmroth2000applying,choi1996design,BDGJNS_IPDPS_2014,buttari2008parallel,cosnard1986parallel,chu1990qr}, but focus has predominantly been on 2D blocked algorithms. 3D algorithms for QR have been proposed~\cite{tiskin2007communication,Solomonik:2017:CPA:3087556.3087561}.

Algorithms~\ref{alg:cqr} and~\ref{alg:cqr2} give pseudocode for the sequential CholeskyQR2 (CQR2) algorithm. It is composed of matrix multiplications and Cholesky factorizations and unlike other QR factorization algorithms, does not require explicit QR factorizations \cite{Fukaya:2014:CSC:2691142.2691147}. Using the building block algorithms explored above, we seek to extend the existing parallel 1D-CQR2 algorithm given in Algorithms~\ref{alg:1dcqr} and~\ref{alg:1dcqr2} to efficiently handle an arbitrary number of rows and columns.

\par{}
\begin{algorithm}
{\algsmall
\caption {$[Q, R] \gets \textbf{CQR}\left(A\right)$}
\label{alg:cqr}
\begin{algorithmic}[1]
\Require{$A$ is $m\times n$}
  \State $W \gets \textbf{Syrk}\left(A\right)$
  \State $R^{T},R^{-T} \gets \textbf{CholInv}\left(W\right)$
  \State $Q \gets \textbf{MM}\left(A, R^{-1}\right)$
\Ensure{$A=QR$, where $Q$ is $m\times n$ with orthonormal columns, $R$ is $n\times n$ upper triangular}
\end{algorithmic}
}
\end{algorithm}
\vspace{0.25cm}
\begin{algorithm}
{\algsmall
\caption {$[Q, R] \gets \textbf{CQR2}\left(A\right)$}
\label{alg:cqr2}
\begin{algorithmic}[1]
\Require{$A$ is $m\times n$}
  \State $Q_{1}, R_{1} \gets \textbf{CQR}\left(A\right)$
  \State $Q, R_{2} \gets \textbf{CQR}\left(Q_{1}\right)$
  \State $R \gets \textbf{MM}(R_{2}, R_{1})$
\Ensure{$A=QR$, where $Q$ is $m\times n$ with orthonormal columns, $R$ is $n\times n$ upper triangular}
\end{algorithmic}
}
\end{algorithm}

\subsection{1D-CholeskyQR2}
\par
\begin{figure}
\centering
\includegraphics[scale=0.65]{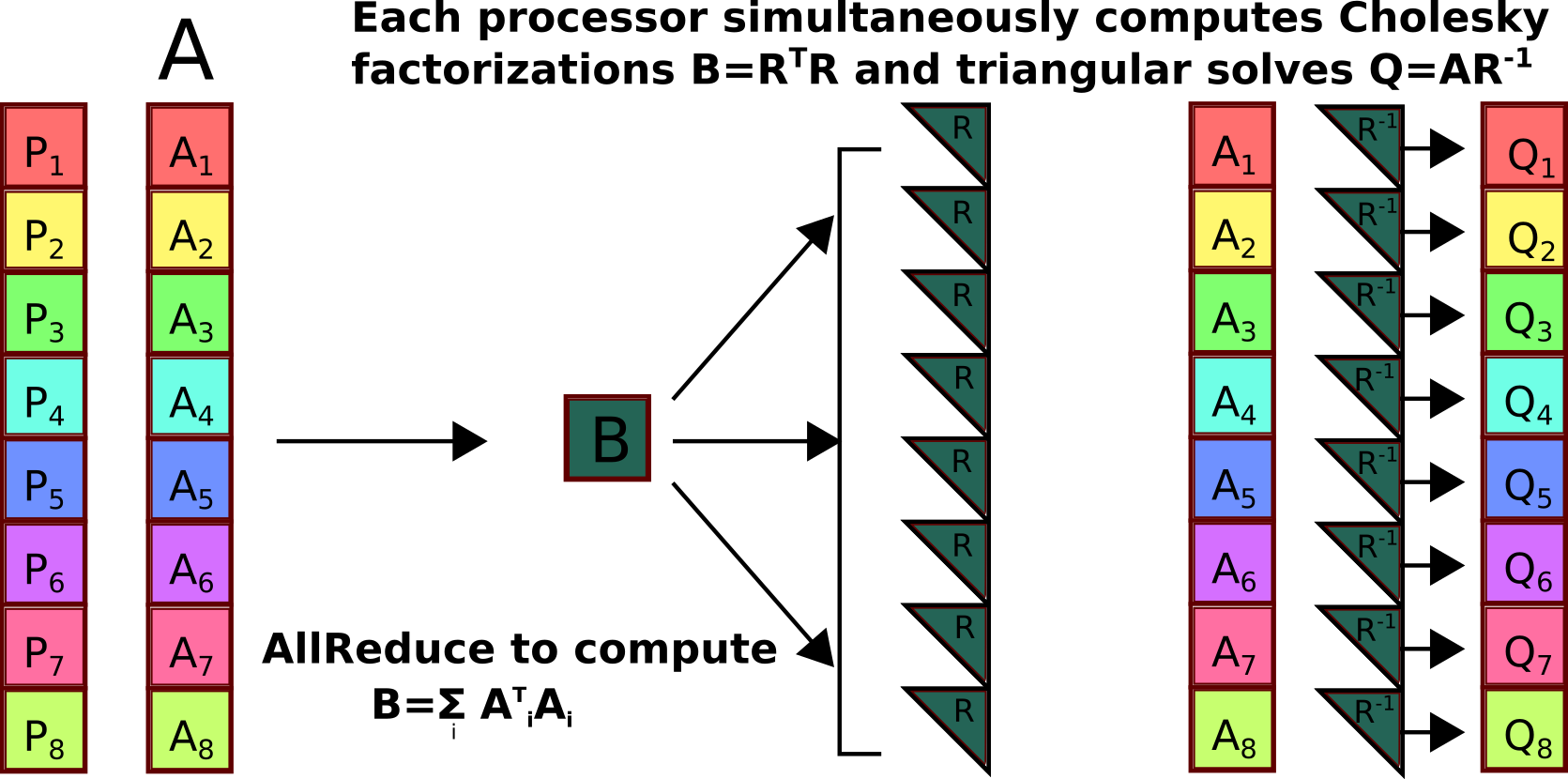}
\caption[]{Illustration of each step in the existing parallel 1D-CQR algorithm.}
\label{fig:CholeskyQR1D}
\end{figure}

{}
The existing parallel 1D-CQR2 algorithm is solved over 1D processor grid $\Pi$ \cite{Fukaya:2014:CSC:2691142.2691147}. It partitions the $m\times n$ matrix $A$ into $P$ rectangular chunks of size $\sfrac{m}{P}\times n$. Each processor performs a sequential symmetric rank-$\sfrac{m}{P}$ update (syrk) with its partition of $A$, resulting in $n\times n$ matrix $\loc{X}=\loc{A}^{T}\loc{A}, \forall p \in [0,P{-}1]$. 1D parallelization allows each processor to perform local matrix multiplication with its initial partition of $A$ and to contribute to the summation across rows using an Allreduce. Each processor performs a sequential Cholesky factorization on the resulting matrix and solves for $R^{-T}$. Because $Q$ is distributed in the same manner as $A$, horizontal communication is not required and each processor can solve for $\loc{Q}$ with $\loc{A}$ and its copy of $R^{-1}$. See Figure \ref{fig:CholeskyQR1D} and Algorithm~\ref{alg:1dcqr} for further details.

\par{}
1D-CQR2 calls 1D-CQR twice to solve for $Q$ as shown in Algorithms~\ref{alg:cqr2} and~\ref{alg:1dcqr2}. Each processor can solve for $R \gets R_{2}R_{1}$ sequentially. This algorithm ensures that $Q$ is distributed the same as $A$ and $R$ is stored on every processor. 

\par{}
In general, 1D-CQR2 can only be applied to extremely overdetermined matrices, where $m > n/P$ and $n$ is small enough to make the Allreduce feasible under given memory constraints. In particular, the algorithm incurs per processor memory footprint and computation costs of $O(n^2)$ and $O(n^3)$, respectively. This algorithm performs well under these conditions \cite{Fukaya:2014:CSC:2691142.2691147}, yet otherwise achieves poor scalability in communication, computation, and memory footprint. We show that CA-CQR2 can scale efficiently when $m > n$.

\par{}
\begin{algorithm}
{\algsmall
\caption {$[\loc{Q}, R] \gets \textbf{1D-CQR}\fleft(\loc{A}, \Pi\fright)$}
\label{alg:1dcqr}
\begin{algorithmic}[1]
\Require{$\Pi$ has $P$ processors arranged in a 1D grid. Each processor owns a (cyclic) blocked partition of $m\times n$ input matrix $A$ given by $\loc{A} \in \mathbb{R}^{\frac{m}{P}\times n}$. Let $X$ be distributed the same as $Q$, $Z$ be distributed the same as $R$.}
  \State $\loc{X} \gets \textbf{Syrk}\fleft(\loc{A}\fright)$ \label{li:alg:1dcqr:syrk}
  \State $\textbf{Allreduce}\fleft(\loc{X}, Z, \Pi\fright)$  \label{li:alg:1dcqr:allred} 
  \State $R^{T},R^{-T} \gets \textbf{CholInv}\fleft(Z\fright)$ \label{li:alg:1dcqr:cinv}
  \State $\loc{Q} \gets \textbf{MM}\fleft(\loc{A}, R^{-1}\fright)$ \label{li:alg:1dcqr:mm}
\Ensure{$A=QR$, where $Q$ is distributed the same as $A$, $R$ is an upper triangular matrix of dimension $n$ owned locally by every processor}
\end{algorithmic}
}
\end{algorithm}

\par{}
\begin{algorithm}
{\algsmall
\caption {$[\loc{Q}, R] \gets \textbf{1D-CQR2}\fleft(\loc{A}, \Pi\fright) $}
\label{alg:1dcqr2}
\begin{algorithmic}[1]
\Require{Same result as Algorithm~\ref{alg:1dcqr}.}
  \State $\loc{X}, W \gets \textbf{1D-CQR}\fleft(\loc{A}, \Pi\fright)$ \label{li:alg:1dcqr2:cqr1}
  \State $\loc{Q}, Z \gets \textbf{1D-CQR}\fleft(X, \Pi \fright)$ \label{li:alg:1dcqr2:cqr2}
  \State $R \gets \textbf{MM}\fleft(Z, W\fright)$ \label{li:alg:1dcqr2:mm}
\Ensure{Same requirements as Algorithm~\ref{alg:1dcqr}.}
\end{algorithmic}
}
\end{algorithm}
See Tables \ref{table:CQR1Dcosts} and \ref{table:CQR21Dcosts} for the costs attained in the 1D-CQR and 1D-CQR2 algorithms, respectively. The overall asymptotic cost is listed in Table \ref{tab:allcosts}.

{\mathsmall
\begin{table}
\renewcommand{\arraystretch}{1.5}
\centering
\makebox[0pt][c]{\parbox{.5\textwidth}{%
\begin{minipage}[b]{.5\hsize}\centering
\begin{tabular}{ l | l }
\# & Cost \\ \hline
\ref{li:alg:1dcqr:syrk} & $\tsyrk{m/P}{n}$ 
 \\ \hline 
\ref{li:alg:1dcqr:allred} & $\tallred{n^2}{P}$ 
 \\ \hline 
\ref{li:alg:1dcqr:cinv} & $\tcholinv{n}$ 
 \\ \hline 
\ref{li:alg:1dcqr:mm} & $\tmm{m/P}{n}{n}$ 
 \\ 
\end{tabular}
\caption{Per-line costs of Algorithm \ref{alg:1dcqr}.}
\label{table:CQR1Dcosts}
\end{minipage}
\hfill
\begin{minipage}[b]{.5\hsize}\centering
\begin{tabular}{ l | l }
\# & Cost \\ \hline
\ref{li:alg:1dcqr2:cqr1} & $\tdcqr{m}{n}{P}$ 
\\ \hline
\ref{li:alg:1dcqr2:cqr2} &  $\tdcqr{m}{n}{P}$ 
\\ \hline
\ref{li:alg:1dcqr2:mm} & $(1/3)n^3 \cdot \gamma$ \\ 
\end{tabular}
\caption{Per-line costs of Algorithm \ref{alg:1dcqr2}.}
\label{table:CQR21Dcosts}
\end{minipage}
}
}%
\end{table}
}

\section{Communication-avoiding CholeskyQR2}
\par{}
Our main goal is to develop a parallelization of Cholesky-QR2 that scales efficiently for rectangular matrices of arbitrary dimensions $m\geq n$. We start with a 3D algorithm best suited when $m=n$.

\subsection{3D-CQR2}
Matrix $A$ is initially distributed across $\Pi[\ALL,\ALL,\pz], \forall \pz \in [0,P^{\sfrac{1}{3}}{-}1]$ and is partitioned into rectangular chunks of size $\sfrac{m}{P^{\sfrac{1}{3}}}\times \sfrac{n}{P^{\sfrac{1}{3}}}$. 
We compute $Z=A^{T}A$ in a similar way to MM3D. Processor $\Pi[\pz,\py,\pz]$ broadcasts $\loc{A}$ along $\Pi[\ALL,\py,\pz]$ as $\loc{W}$. Local matrix multiplication with $\loc{W}^{T}$ and $\loc{A}$ produces $\loc{X}$. A reduction along $\Pi[\px,\ALL,\pz]$ onto root $\pz$ sums each $\loc{X}$ along the corresponding subgrid, resulting in a unique partition of $X$ scattered across $\Pi$. A broadcast along $\Pi[\px,\py,\ALL], \forall \px,\py \in [0,P^{\sfrac{1}{3}}{-}1]$ fills in the missing partitions and ensures that each $\Pi[\ALL,\ALL,\pz], \forall \pz \in [0,P^{\sfrac{1}{3}}{-}1]$ owns a copy of $Z$.

\par{}
With $Z$ distributed the same as $A$, CFR3D and MM3D can solve for $R^{T}$ and $Q$, respectively. An alternate strategy involves computing triangular inverted blocks of dimension $n_{o}$ and solving for $Q$ with multiple instances of MM3D. This strategy can lower the computational cost by nearly a factor of 2 when $n_{o}=n/2$, incurring close to a $2x$ increase in synchronization cost. As $n_{o}$ decreases by a factor of $2$, the synchronization cost increases logarithmically and the decrease in computational cost remains nearly the same.
This algorithm attains a communication cost (see Table~\ref{tab:allcosts}) equivalent to the lower bounds of $LU$ and matrix multiplication \cite{irony_mm_lb04,Solomonik:2017:TSC:2965648.2897188}.

\subsection{CA-CQR2}
\par{}
Tunable processor grids have several advantages over static grids when factoring rectangular matrices. They can match the shape of the matrix and tune themselves to optimize certain parameters such as memory size and horizontal communication. These costs can be shown to interpolate between known algorithms on specific grids. Skinny matrices cannot take full advantage of the resources provided by a 3D grid, while square matrices overload the resource capacity that skinny rectangular grids provide.
\par{}
The CA-CQR2 algorithm can be seen as a generalization of 1D-CQR2 and 3D-CQR2. We define a $c\times d\times c$ rectangular processor grid $\Pi$ that partitions the $m\times n$ matrix $A$ into rectangular blocks of size $\frac{m}{d}\times \frac{n}{c}$. CA-CQR2 effectively utilizes its tunable grid by performing $d/c$ simultaneous instances of CFR3D on cubic grid partitions of dimension $c$. This allows each grid partition to avoid further communication with other partitions because each has the necessary data to compute the final step $Q=AR^{-1}$.
\par{}
As in 3D-CQR, processor $\Pi[\pz,\py,\pz]$ broadcasts $\loc{A}$ to $\Pi[\ALL,\py,\pz], \forall \py \in [0,d{-}1], \pz \in [0,c{-}1]$ as $\loc{W}$, after which a local matrix multiplication $\loc{X} \gets \loc{W}^{T}\loc A$ is performed. A few extra steps are necessary in order for $d/c$ cubic partitions of $\Pi$ to own the same matrix $Z=A^{T}A$. First, we subdivide $\Pi$ along dimension $y$ into $d/c$ contiguous groups of size $c$. 
Each group then participates in a reduction onto root processor $\Pi[\px,\py,\pz], \forall \px,\pz \in [0,c{-}1], \forall \py$ such that $\py \bmod c = \pz$. To complete the linear combination, we subdivide $\Pi$ along dimension $y$ into $d/c$ groups of size $c$, where each processor belonging to the same group is a step size $c$ away. An Allreduce is performed on this subcommunicator resulting in every processor owning a single correct submatrix of $B=A^{T}A$. A final broadcast from root $\Pi[\px,\py,\pz], \forall \px,\pz \in [0,c{-}1], \forall \py$ such that $\py \bmod c = \pz$ along $\Pi[\px,\py,\ALL], \forall \px \in [0,c{-}1], \py \in [0,d{-}1]$ ensures that each copy of $B$ is distributed over all $c\times c$ processor subgrids.

\par{}
With $Z$ distributed as described above, $\frac{d}{c}$ simultaneous instances of CFR3D and MM3D are performed over $\Pi_\text{subcube}$ to obtain $Q$. CA-CQR2 requires calling CA-CQR twice and performing a final MM3D to solve for $R$.

\par
\begin{figure}
\centering
\includegraphics[scale=0.44]{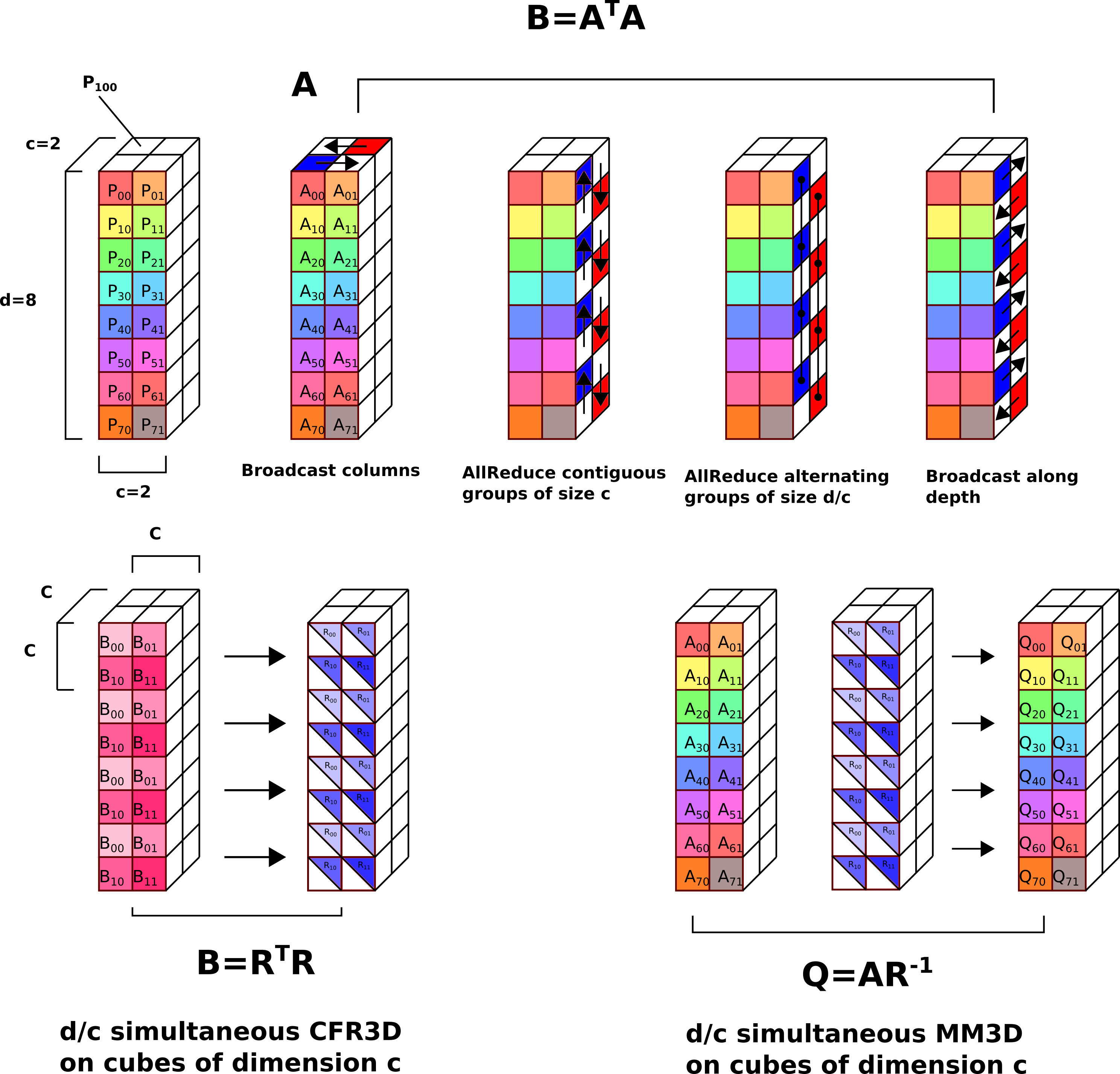}
\caption[]{Illustration of the steps required to perform CQR over a tunable processor grid.}
\label{fig:CholeskyQRTune}
\end{figure}

{}
CA-CQR2 combines elements of both the 1D-CQR2 and 3D-CQR2 algorithms in order to most efficiently span the grid range $c \in [1,P^{\sfrac{1}{3}}]$.
Further details are provided in Figure \ref{fig:CholeskyQRTune}, pseudocode presented in Algorithms~\ref{alg:cacqr} and~\ref{alg:cacqr2}, and the cost analysis below.
Tables \ref{table:CQRTunablecosts} and \ref{table:CQR2Tunablecosts} present the costs attain in the CA-CQR and CA-CQR2 algorithms, respectively.

\par{}
\begin{algorithm}
{\algsmall
\caption {$[\loc Q, \loc R] \gets \textbf{CA-CQR}\left(\loc A, \Pi\right)$}
\label{alg:cacqr}
\begin{algorithmic}[1]
\Require{$\Pi$ has $P$ processors arranged in a tunable grid of size $c\times d\times c$ for any integer $c$ in range $[0, P^{\frac{1}{3}}{-}1]$. $A$ is $m\times n$ and is replicated on $\Pi[\ALL,\ALL,\pz], \forall \pz \in [0,c{-}1]$. Each processor $\Pi[\px,\py,\pz]$ owns a (cyclic) blocked partition of $A$ given by $\loc A\in\mathbb{R}^{\frac{m}{d}\times \frac{n}{c}}$. Let $W$, $X$, $Y$, and $Z$ be temporary arrays distributed the same as $A$.}
  \State \label{li:alg:cacqr:bc1}   $\textbf{Bcast}\fleft(\loc{A}, \loc{W}, \pz, \Pi[\ALL,\py,\pz]\fright)$
  \State \label{li:alg:cacqr:mm1}   $\loc{X} \gets \textbf{MM}\fleft(\loc{W}^{T}, \loc{A}\fright)$
  \State \label{li:alg:cacqr:red1}  $\textbf{Reduce}\fleft(\loc X, \loc{Y}, \pz, \Pi[\px,c\lfloor\sfrac{\py}{c}\rfloor:c\lceil\sfrac{\py}{c}\rceil-1,\pz]\fright)$
  \State \label{li:alg:cacqr:allred1} $\textbf{Allreduce}\fleft(\loc{Y}, \loc{Z}, \Pi[\px,\py\mod c:c:d-1,\pz]\fright)$
  \State \label{li:alg:cacqr:bc2}     $\textbf{Bcast}\fleft(\loc{Z}, \loc{Z}, \py\mod c, \Pi[\px,\py,\ALL]\fright)$
  \State \label{li:alg:cacqr:def} Define $\Pi_\text{subcube}[\px,\py,\pz] \gets \Pi[\px,c\lfloor\sfrac{\py}{c}\rfloor:c\lceil\sfrac{\py}{c}\rceil-1,\pz]$
  \State \label{li:alg:cacqr:cfr}   $\loc{R^{T}}, \loc{R^{-T}} \gets \CFRDDD\fleft(\loc{Z}, \Pi_\text{subcube}\fright)$
  \State \label{li:alg:cacqr:mm3d}  $\loc Q \gets \textbf{MM3D}\fleft(\loc A, \loc{R^{-1}}, \Pi_\text{subcube}\fright)$
\Ensure{$A=QR$, where $Q$ and $R$ are distributed the same as $A$. $Q$ is $m\times n$ orthonormal and $R$ is an upper triangular matrix of dimension $n$.}
\end{algorithmic}
}
\end{algorithm}
\begin{algorithm}
{\algsmall
\caption {$[\loc Q, \loc R] \gets \textbf{CA-CQR2}\left(\loc A, \Pi\right)$}
\label{alg:cacqr2}
\begin{algorithmic}[1]
\Require{Same requirements as Algorithm~\ref{alg:cacqr}.}
  \State $\loc X, \loc Y \gets \textbf{CA-CQR}\left(\loc A, \Pi\right)$ \label{li:alg:cacqr2:cacqr1}
  \State $\loc Q, \loc Z \gets \textbf{CA-CQR}\left(\loc X, \Pi\right)$ \label{li:alg:cacqr2:cacqr2}
  \State Define $\Pi_\text{subcube}[\px,\py,\pz] \gets \Pi[\px,c\lfloor\sfrac{\py}{c}\rfloor:c\lceil\sfrac{\py}{c}\rceil,\pz]$
  \State $\loc R \gets \textbf{MM3D}\left(\loc Z,\loc  Y, \Pi_\text{subcube}\right)$ \label{li:alg:cacqr2:mm1}
\Ensure{Same requirements as Algorithm~\ref{alg:cacqr}.}
\end{algorithmic}
}
\end{algorithm}

\par{}
{\mathsmall
\begin{table}
\renewcommand{\arraystretch}{1.5}
\centering
\makebox[0pt][c]{\parbox{.5\textwidth}{%
\begin{minipage}[b]{.5\hsize}\centering
\begin{tabular}{ l | l }
\hline
\# & Cost \\ \hline
\ref{li:alg:cacqr:bc1}   & $\tbcast{\frac{mn}{dc}}{c}$ 
 \\ \hline 
\ref{li:alg:cacqr:mm1}   & $\tmm{\sfrac nc}{\sfrac md}{\sfrac nc}$ 
 \\ \hline 
\ref{li:alg:cacqr:red1}  & $\tred{\sfrac{n^2}{c^2}}{c}$ 
 \\ \hline 
\ref{li:alg:cacqr:allred1}  & $\tallred{\sfrac{n^2}{c^2}}{d/c}$ 
 \\ \hline 
\ref{li:alg:cacqr:bc2}      & $\tbcast{\frac{mn}{dc}}{c}$ 
 \\ \hline 
\ref{li:alg:cacqr:cfr}  & $\tcfrddd{n}{c^3}$ 
 \\ \hline 
\ref{li:alg:cacqr:mm3d} & $\tmmddd{m}{n}{n}{c^3}$ 
 \\ \hline 
\end{tabular}
\caption{Per-line costs of Algorithm \ref{alg:cacqr}.}
\label{table:CQRTunablecosts}
\end{minipage}
\hfill
\begin{minipage}[b]{.5\hsize}\centering
\begin{tabular}{ l | l }
\hline
\# & Cost \\ \hline
\ref{li:alg:cacqr2:cacqr1} & $\tcacqr mncd$ 
 \\ \hline 
\ref{li:alg:cacqr2:cacqr2} &  $\tcacqr mncd$ 
 \\ \hline 
\ref{li:alg:cacqr2:mm1} & $\tmmddd{n}{n}{n}{c^3}$ 
 \\ \hline 
\end{tabular}
\caption{Per-line costs of Algorithm \ref{alg:cacqr2}.}
\label{table:CQR2Tunablecosts}
\end{minipage}
}
}%
\end{table}
}

\par{}
The overall asymptotic cost (also listed in Table \ref{tab:allcosts}) is
{\small 
\begin{equation*}
\begin{aligned}
T^{\alpha-\beta}_{\text{CA-CQR2}}\left(m,n,c,d\right) = \mathcal{O}\bigg(&c^{2}\log P \cdot \alpha + \Big(\frac{mn\delta(c)}{dc} +\frac{n^2}{c^2}\Big)\cdot \beta \\
&+ \Big(\frac{mn^2}{c^{2}d} + \frac{n^3}{c^3}\Big) \cdot \gamma\bigg).
\end{aligned}
\end{equation*}
}%
The overall memory footprint is $mn/dc + n^{2}/c^{2}$.

\par{}
With appropriate choices of $c$ and $d$, the costs attained by CA-CQR2 interpolate between the costs of 1D-CQR2 and 3D-CQR2. Optimal communication can be attained by ensuring that the grid perfectly fits the dimensions of $A$, or that the dimensions of the grid are proportional to the dimensions of the matrix. We derive the cost for the optimal ratio $\frac{m}{d}=\frac{n}{c}$ (last row of Table~\ref{tab:allcosts}).
Both the memory footprint and communication cost of this algorithm are $(\sfrac{mn^2}{P})^{\sfrac{2}{3}}$.

\begin{figure*}[t]
\captionsetup[subfigure]{labelformat=empty}
\centering
\subfigure{
    \includegraphics[width=0.48\textwidth]{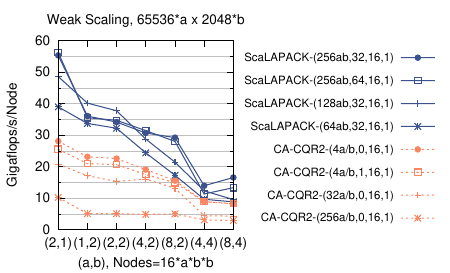}
}
\subfigure{
    \includegraphics[width=0.48\textwidth]{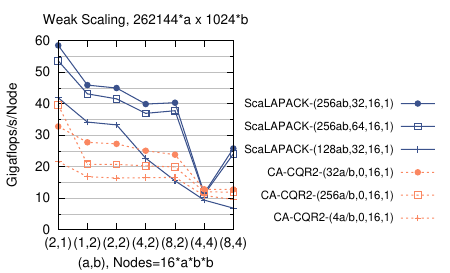}
}
\end{figure*}

\begin{figure}[ht]
\vspace*{-.3in}
\centering
    \includegraphics[width=0.48\textwidth]{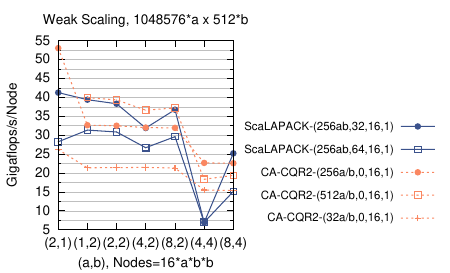}
\caption{(a,b,c) Weak scaling performance Blue Waters.}
\label{fig:bw_ws_plot}
\end{figure}

\section{Evaluation}
\par{}
We study the performance of our implementation of the CA-CQR2 algorithm, with respect to ScaLAPACK's PGEQRF routine on two supercomputers: Blue Waters and Stampede2\footnote{We used our own build of ScaLAPACK 2.0.2 on Stampede2, after finding that the existing MKL implementation stalled during execution of QR with most processor grids when using 131072 or more MPI processes. We did not observe a noticeable difference in performance between the two versions.}.
All variants of CholeskyQR2, including CA-CQR2, perform $4mn^{2}+\frac{5n^{3}}{3}$ flops along its critical path, while ScaLAPACK's PGEQRF uses Householder QR and performs $2mn^{2}-\frac{2}{3}n^{3}$ such flops.
We show that while CA-CQR2 is somewhat slower than ScaLAPACK on Blue Waters and when using few nodes of Stampede2, it outperforms PGEQRF when using a large number of nodes of Stampede2 despite this increase in computation. 
A key difference between Stampede2 and Blue Waters is the ratio of peak floating performance to network communication bandwidth.
Specifically, the ratio of peak flops to injection bandwidth is roughly 8X higher on Stampede2.
This difference also reflects an architectural trend of flop rates increasing faster than bandwidth in high-performance computing architectures.
CA-CQR2 is thus better-fit for massively-parallel execution on newer architectures as it reduces communication at the cost of computation, yielding a higher arithmetic intensity.
\par{}
Our scaling studies also illustrate the effects parameters $d$ and $c$ have on computation cost, communication cost, synchronization cost, and performance as a whole.
In particular, the scaling results on Stampede2 provide the first experimental evidence that performance improvements and superior scaling can be attained by increasing the memory footprint to reduce communication for QR factorization.
Specifically, the parameter $c$ determines the memory footprint overhead; the more replication being used ($c$), the larger the expected communication improvement ($\sqrt{c}$) over 2D algorithms.
\subsection{Implementation}
\par{}
Our CA-CQR2 implemetation uses C++ and MPI~\cite{Gropp:1994:UMP:207387} for all collective and point-to-point communication routines.
Sequential matrix multiplication and matrix factorization routine use BLAS~\cite{lawson1979basic} and LAPACK~\cite{LAPACK}.
The implementation does not exploit overlap in computation and communication and used a cyclic data distribution among processors.

\subsection{Architectural Setup}
\par{}
We use both the Stampede2 supercomputer at Texas Advanced Computing Center (TACC)\cite{Stanzione:2017:SEX:3093338.3093385} and the Blue Waters supercomputer at the National Center for Supercomputing Applications (NCSA).
Stampede2 consists of 4200 Intel Knights Landing (KNL) compute nodes (each capable of a performance rate over 3 Teraflops/s) connected by an Intel Omni-Path (OPA) network with a fat-tree topology (achieving an injection bandwidth of 12.5 GB/sec).
Each KNL compute node provides 68 cores with 4 hardware threads per core.
Both our implementation and ScaLAPACK use he Intel/17.0.4 environment with MPI version impi/17.0.3.
The configuration of CA-CQR2 uses optimization flag -03 and full 512-bit vectorization via flag -xMIC-AVX512.

Blue Waters is a Cray XE/XK hybrid machine composed of AMD Interlagos processors and NVIDIA GK110 accelerators connected by a Cray Gemini 3D torus interconnect (achieving an injection bandwidth of 9.6 GB/s).
We utilize XE compute nodes which each hold 16 floating-point Bulldozer core units and can achieve a performance rate of 313 Gigaflops/s.
On Blue Waters, we use the PrgEnv-gnu/5.2.82 module with MPI version cray-mpich/7.5.0 and gcc/4.9.3 compiler.
The optimization flags include -O3, -ffast-math, -funroll-loops, -ftree-vectorize, and -std=gnu++11.
We utilize Cray's LibSci implementations of BLAS/LAPACK routines via module cray-libsci/16.11.1.

\subsection{Experimental Setup}
\par{}
For both strong and weak scaling tests, we generate random matrices.
Each data-point displayed in figures corresponds to the median execution time out of five iterations.
The performance variation (due to noise) did not seem significant on both Blue Waters and Stampede2.

We test strong scaling by keeping the $m\times n$ matrix size constant while increasing the number of nodes $N$, the CA-CQR2 grid dimension $d$, and the row-dimension of ScaLAPACK's 2D processor-grid $p_r$ by a factor of $2$.
We test weak scaling by keeping both the dimensions of local matrices and the leading-order flop cost constant while increasing both matrix dimensions, processor grid dimensions, and $N$.
We alternate between two scaling progressions:
\begin{enumerate}
\item increase $m$, $d$, and $p_r$ by a factor of $2$ while leaving $n$, CA-CQR2 grid dimension $c$, and the column-dimension of ScaLAPACK's processor grid $p_c$ constant,
\item decrease $m$ and $d$ by a factor of $2$, increase $n$ and $c$ by a factor of $2$, while leaving $p_r$ constant.
\end{enumerate}
The first progression is employed 3x as often as the second.
\par{}
We test CA-CQR2 against ScaLAPACK's PGEQRF routine over a range of matrix sizes, block sizes, processor grid dimensions, processes per node ($ppn$), and thread counts per MPI process ($tpr$).
In the weak scaling plots, each CA-CQR2 variant is characterized by a tuple $(d/c,InverseDepth,ppn,tpr)$, where $a$ and $b$ are defined on each plot.
In the strong scaling plots, each CA-CQR2 is characterized by a tuple $(d,c,InverseDepth,ppn,tpr)$.
$InverseDepth$ determines the last recursive level at which the triangular inverse factor is computed. $InverseDepth$ can range between $0$ and $c/4$.
In both scaling plots, each PGEQRF variant is characterized by a tuple $(p_r,BlockSize,ppn,tpr)$.
\par{}
We vary the $ppn$ and $tpr$ in all tests, choosing to include the most performant variants among ($ppn,tpr$)=(8,8),(16,4),(16,8),(32,2),(32,4),(64,1),(64,2) on Stampede2, and ($ppn,tpr$)=(16,1),(16,2) on Blue Waters.
Gigaflop/s is calculated for both CA-CQR2 and ScaLAPACK, by dividing $2mn^{2}-\frac{2}{3}n^{3}$ by the execution time (ignoring the extra computation done by CA-CQR2).
Therefore, CA-CQR2 achieves a 2x to 4x greater percentage of peak performance than the plots depict.

\subsection{Weak Scaling}
\begin{figure*}[t]
\captionsetup[subfigure]{labelformat=empty}
\centering
\subfigure{
    \includegraphics[width=0.48\textwidth]{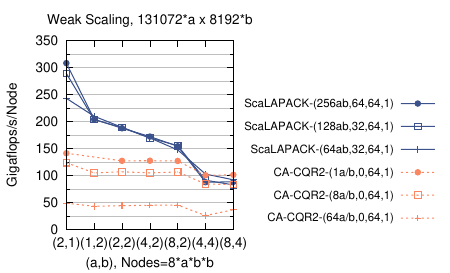}
    \label{fig:stampede2_ws_plot_1}
}
\subfigure{
    \includegraphics[width=0.48\textwidth]{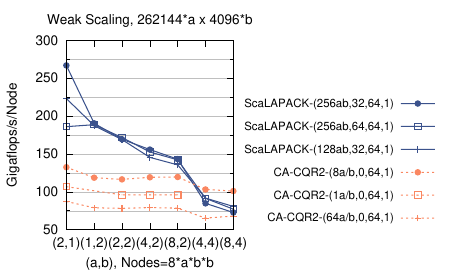}
    \label{fig:stampede2_ws_plot_2}
}
\subfigure{
    \includegraphics[width=0.48\textwidth]{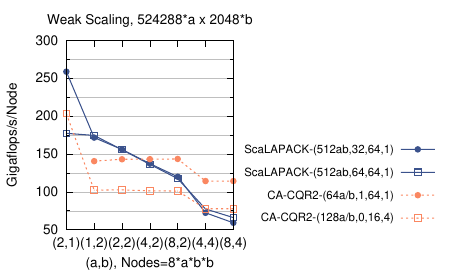}
    \label{fig:stampede2_ws_plot_3}
}
\subfigure{
    \includegraphics[width=0.48\textwidth]{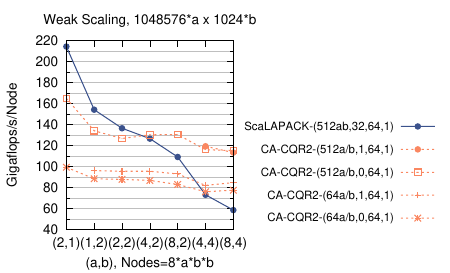}
    \label{fig:stampede2_ws_plot_4}
}
\caption{(a,b,c,d) Weak scaling performance Stampede2.}
\label{fig:stampede2_ws_plot}
\end{figure*}

We begin by analyzing results for weak scaling (keeping work per processor constant) on Blue Waters. Figure~\ref{fig:bw_ws_plot} 
demonstrates the performance of our implementation of CA-CQR2 and ScaLAPACK for differently dilated processor grids and matrices with an increasing (across plots) ratio of rows and columns. Within each variant of CA-CQR2, for weak scaling, both $m/d$ to $n/c$ are kept constant, so communciation and computation cost stay constant, giving the increasing synchronization cost an increasingly dominant effect on performance. We confirm that 1D-CQR2 and CA-CQR2 variants with small $c$ are not suited for matrices with a large number of columns. 
For example, for Figure 4(c), when using $N=32$ nodes, as $c$ increases from $c=1$ to $c=2$ (so the first parameter in the legend label, $d/c$, decreases from $512a/b$ to $64a/b$), we observe a factor of 2x decrease in execution time, since while the synchronization cost increases by roughly 4x, the communication cost decreases by roughly $\sqrt{2}$x and the computation cost decreases by roughly 4x (suggesting execution time is dominated by a mix of computation and communication costs). 
The sensitivity to computation cost can also be observed from other Blue Waters performance data and in part explains CA-CQR2's poor performance relative to ScaLAPACK's QR on this architecture.
\par{}
As the ratio of the number of rows to columns increases from Figure 4(a)
to Figure 4(c),
CA-CQR2 shows increasingly better performance relative to PGEQRF, which may be attributed to the decrease in computational cost overhead ($O(n^3)$ sequential cost term associated with Cholesky factorizations) relative to Householder QR factorization. Indeed, both the highest absolute performance and highest performance relative to ScaLAPACK's QR on Blue Waters is achieved for matrices with a relatively smaller number of columns due to the increase in critical path cost (algorithm depth / latency cost) associated with increasing $n$.
\par{}
Weak scaling results on Stampede2 exhibit similar trends among CA-CQR2 algorithm variants, yet give performance improvements over ScaLAPACK's QR at a range of node counts determined by the shape of the matrix. Figure~\ref{fig:stampede2_ws_plot} demonstrates that as the ratio of rows to columns increases across plots (and associated flop cost overhead decreases), CA-CQR2's $P^{1/6}$ reduction in communication can influence performance at increasingly smaller node counts. CA-CQR2 improves ScaLAPACK's QR performance at 1024 nodes by factors of 1.1x in Figure \ref{fig:stampede2_ws_plot_1} with $c=32$, 1.3x in Figure \ref{fig:stampede2_ws_plot_2} with $c=16$, 1.7x in Figure \ref{fig:stampede2_ws_plot_3} with $c=8$, and  1.9x in Figure \ref{fig:stampede2_ws_plot_4} with $c=4$. The most performant variants of each algorithm are compared directly in Figure \ref{fig:BestWSvariants}.
\par{}
The overhead of synchronization is less prevalent on Stampede2 than Blue Waters, as evidenced by the relatively constant performance among variants.
Matrices with more columns (Figures \ref{fig:stampede2_ws_plot_1} and \ref{fig:stampede2_ws_plot_2}) exhibit the best parallel scaling, due to a higher computation cost, since $O(mn^2)$ grows quadratically with $n$. By contrast, tall-and-skinny matrices (Figures \ref{fig:stampede2_ws_plot_3} and \ref{fig:stampede2_ws_plot_4}) are more sensitive to the increasing synchronization cost overhead.
\par{}
The computation cost overheads of CA-CQR2 at 1024 nodes for all matrix variants in Figure~\ref{fig:stampede2_ws_plot} are all roughly a factor of 2x, whereas the algorithm's communication cost reduction relative to ScaLAPACK's QR at 1024 nodes is $4\sqrt{2}$x, $4$x, $2\sqrt{2}$x, and $1$x, respectively. Despite the constant flop cost overhead, CA-CQR2 achieves better performance at increasingly smaller node counts for matrices with increasingly higher row-to-column ratios that benefit from a smaller $c$. Its asymptotically higher synchronization costs associated with larger $c$ can play as significant a role as its higher computational costs in determining when this algorithm is most beneficial on Stampede2.

\subsection{Strong Scaling}
\begin{figure*}[t]
\captionsetup[subfigure]{labelformat=empty}
\centering
\subfigure{
    \includegraphics[width=0.48\textwidth]{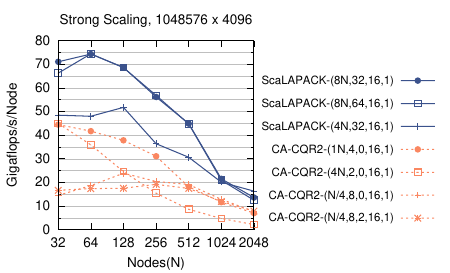}
    \label{fig:bw_ss_plot_1}
}
\subfigure{
    \includegraphics[width=0.48\textwidth]{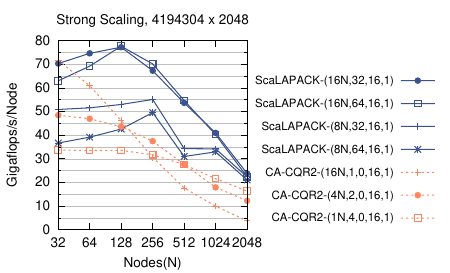}
    \label{fig:bw_ss_plot_2}
}
\label{fig:bw_ss_plot}
\caption{(a,b) Strong scaling performance on Blue Waters.}
\end{figure*}

\begin{figure*}[t]
\captionsetup[subfigure]{labelformat=empty}
\centering
\subfigure{
    \includegraphics[width=0.48\textwidth]{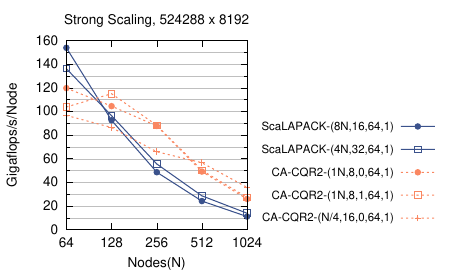}
    \label{fig:stampede2_ss_plot_1}
}
\subfigure{
    \includegraphics[width=0.48\textwidth]{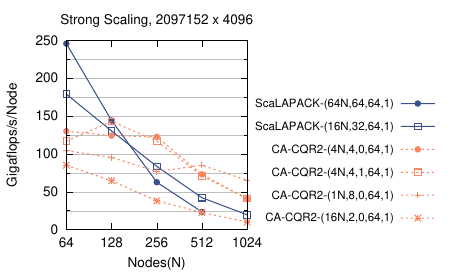}
    \label{fig:stampede2_ss_plot_2}
}
\subfigure{
    \includegraphics[width=0.48\textwidth]{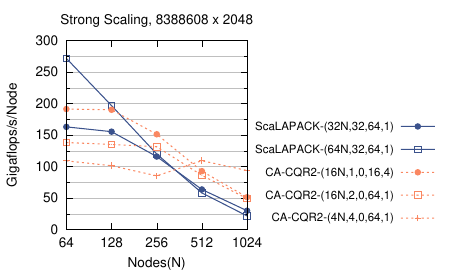}
    \label{fig:stampede2_ss_plot_3}
}
\subfigure{
    \includegraphics[width=0.48\textwidth]{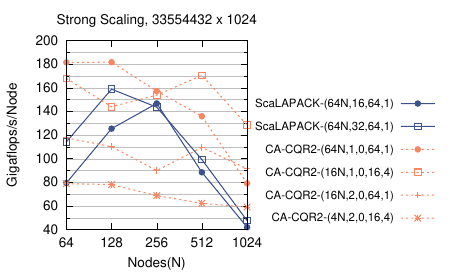}
    \label{fig:stampede2_ss_plot_4}
}
\caption{(a,b,c,d) Strong scaling performance on Stampede2.}
\end{figure*}

We first consider strong scaling on Blue Waters. We reference each CA-CQR2 processor grid by the parameter $c$, which controls the size of two of its dimensions. In Figure \ref{fig:bw_ss_plot_2}, grid $c=1$ attains the best absolute performance at the smallest node count among CA-CQR2 variants. This can be attributed to $mn/dc \gg n^{2}/c^{2}$ and $mn^{2}/dc^{2} \gg n^{3}/c^{3}$, communication and flop cost measures, respectively. In general, processor grids with smaller $c$ will continue to achieve higher performance in strong scaling studies as long as  these conditions are met. As the node count increases, the cost terms dependent solely on the non-scaling parameters $n,c$ start to dominate and the grids with larger $c$ achieve higher performance and better scalability. In Figure \ref{fig:bw_ss_plot_2}, this first crossover point occurs at $N=256$ between grids $c=1$ and $c=2$. The second crossover point occurs at $N=512$ between $c=2$ and $c=4$. At $N=2048$, grid $c=4$ achieves the highest performance because its communication cost and flop cost is smallest. This configuration's higher synchronization cost at $N=2048$ (and all node counts) prevents a more drastic performance dropoff. A crossover point between $c=2$ and $c=4$ is reached immediately in Figure \ref{fig:bw_ss_plot_1} due to a smaller ratio $m/n$. These results signify that matrices with a smaller ratio $m/n$ must employ grids with larger $c$ to reduce overhead.
\par{}
Although ScaLAPACK's QR achieves higher performance at the lower node counts, at $N=2048$, this performance difference is small as our best CA-CQR2 variants ($c=8$ in Figure \ref{fig:bw_ss_plot_1} and $c=4$ in Figure \ref{fig:bw_ss_plot_2}) have scaled more efficiently. At higher node counts, the asymptotic communication improvement achieved by CA-CQR2 is expected to be of greater benefit.
\par{}
The strong scaling results in Figures \ref{fig:stampede2_ss_plot_1}, \ref{fig:stampede2_ss_plot_2}, \ref{fig:stampede2_ss_plot_3}, and \ref{fig:stampede2_ss_plot_4} illustrate that CA-CQR2 achieves both higher absolute performance and better scaling than ScaLAPACK's QR across a range of matrix sizes. The scaling characteristics of grids with largest and smallest $c$ are similar to those of Blue Waters. In all but Figure \ref{fig:stampede2_ss_plot_4}, the processor grids with largest $c$ initially suffer from synchronization overhead, yet attain the highest performance at $N=1024$ by performing the least communication and computation. The grids with smallest $c$ exhibit these characteristics in reverse. As discussed above, the architectural differences between the two machines strengthens the impact of our algorithm's asymptotic communication reduction by diminishing the performance impact of the inherent work overhead in CQR2 algorithms. CA-CQR2 improves ScaLAPACK's QR performance at 1024 nodes by factors of 2.6x in Figure \ref{fig:stampede2_ss_plot_1} with $c=8$, 3.3x in Figure \ref{fig:stampede2_ss_plot_2} with $c=4$, 3.1x in Figure \ref{fig:stampede2_ss_plot_3} with $c=4$, and  2.7x in Figure \ref{fig:stampede2_ss_plot_4} with $c=1$.
The most performant variants of each algorithm are compared directly in Figure \ref{fig:BestSSvariants}.
Overall, we observe that CA-CQR2 can outperform optimized existing QR library routines, especially for strong scaling on architectures with larger compute power relative to communication bandwidth, a key distinguishing factor between Stampede2 and Blue Waters (and a segment of a more general trend).

\section{Conclusion}
\par{}
We have developed an algorithm that efficiently extends CholeskyQR2 (CQR2) to rectangular matrices. 
Our analysis provides new insights into the communication complexity of CQR2 and the performance potential of parallel 3D QR algorithms.
Through the use of a tunable processor grid, CQR2 has been generalized to a parallel algorithm equipped to efficiently factorize a matrix of any dimensions via an appropriate 3D algorithm variant. 
Its simplicity and asymptotically optimal communication complexity gives CA-CQR2 promising potential as a parallel QR factorization algorithm. 
In comparison to the ScaLAPACK PGEQRF routine on Stampede2, our implementation achieves better strong and weak scaling on large node counts.
\par{}
One next step in this line of work would be to 
develop a CA-CQR2 algorithm that operates on subpanels to reduce computation cost overhead is also of interest for near-square matrices.
Additionally, while minimal modifications are necessary to implement shifted Cholesky-QR\cite{2018arXiv180911085F}, a performance evaluation of this variant would provide insight to the potential benefits of this unconditionably stable algorithm at scale.

\section*{Acknowledgments}
\par{}
The first author would like to acknowledge the Department of Energy (DOE) and Krell Institute for support via the DOE Computational Science Graduate Fellowship (Grant number DE-SC0019323). 
This work used the Extreme Science and Engineering Discovery Environment (XSEDE), which is supported by National Science Foundation grant number ACI-1548562. Via XSEDE, the authors made use of the TACC Stampede2 supercomputer (allocation TG-CCR180006). This research is part of the Blue Waters sustained-petascale computing project, which is supported by the National Science Foundation (awards OCI-0725070 and ACI-1238993) and the state of Illinois. Blue Waters is a joint effort of the University of Illinois at Urbana-Champaign and its National Center for Supercomputing Applications.
The authors would also like to acknowledge ALCF for providing HPC resources for preliminary benchmarking.